 \documentclass[pra,twocolumn,nofootinbib,eqsecnum,floatfix]{revtex4}
\usepackage{bm}
\usepackage{amsmath}
\usepackage{graphics, graphicx}
\numberwithin{equation}{section}
\renewcommand{\theequation}{\arabic{section}.\arabic{equation}}

\newcommand{\ltwid}{\mathrel{\raise.3ex\hbox{$<$\kern-.75em\lower1ex\hbox{$\sim$}}}}
\newcommand{\gtwid}{\mathrel{\raise.3ex\hbox{$>$\kern-.75em\lower1ex\hbox{$\sim$}}}}

\def\Re{{\rm Re}}

\def\re{\text{Re}}
\def\DH{\text{DH~}}
\def\EP{\text{EP~}}
\def\cm{\text{cm}}
\def\range{$[0,1]$}
\def\dec{{\rm dec}}

\begin{document}

\title{Quantum Mechanics with  Extended Probabilities}

\author{James B.~Hartle}
\email{hartle@physics.ucsb.edu}

\affiliation{Department of Physics,
 University of California,
 Santa Barbara, CA 93106-9530}

\date{\today}

\begin{abstract}
The quantum mechanics of closed systems such as the universe is formulated using an extension of familiar probability theory that incorporates negative probabilities. Probabilities must be positive for sets of alternative histories that are the basis of fair settleable bets.  However, in quantum mechanics there are  sets of alternative histories that can be described  but which cannot be the basis for fair settleable bets. Members of such sets can be assigned extended probabilities that are sometimes negative. A prescription for extended probabilities is introduced that assigns extended probabilities to all histories that can be described, fine grained or coarse grained, members of decoherent sets or not.  All probability sum rules are satisfied exactly. Sets of histories that are recorded to sufficient precision are the basis of settleable bets. This formulation is compared with the decoherent (consistent) histories formulation of quantum theory.  Prospects are discussed  for using this formulation to provide testable alternatives to quantum theory or further generalizations of it.


\vskip 1in
\end{abstract}


\maketitle

\section{Introduction}
\label{sec1}

Probabilities can be understood as instructions for making fair bets \cite{deF37}. From this point of view, the probabilities supplied by well confirmed physical theories, such as classical, statistical, and quantum mechanics, provide secure ways of betting on the regularities exhibited by our universe. Such bets are effectively made every day in the development of technology. Starting from this perspective, this paper extends usual probability theory and generalizes the modern quantum mechanics of closed systems so that extended probabilities are assigned to every history the system can exhibit. 

No bet is complete without specifying the  means to settle it.  Typically this is by appeal to a {\it record} of which of the possible alternatives occurred. 
Usual probability theory implicitly assumes there is a way of settleing a bet on the alternatives to which probablities are assigned.   

The  rules of probability theory follow from a requirement that it not be possible to arrange unfair settleable  bets in which one party is sure to lose \cite{deF37,Cavup}. In particular, probabilities must be positive to prevent this as we review in Section \ref{sec2}. 

But in quantum theory there are alternatives which  can be described but that are not the basis for settleable bets. The two-slit experiment in Figure 1 provides an example in the context of the approximate (Copenhagen) quantum mechanics of measured subsystems.  An electron starts at a source, passes through a screen with two slits, and is detected at a point $y$ on a further screen. Consider the two alternative histories distinguished by whether the electron went through the upper slit or the lower slit to arrive at a given point $y$. 

If a measurement determines which slit the electron passed through, quantum mechanics provides probabilities for a bet  on which occurred. A record of the measurement outcome can be used to settle the bet. If no measurement is carried out, the alternative histories going through the upper and lower slit can still  be described.  But, because of quantum interference, there can be no record of which occurred. 
A bet on these alternative histories is not settleable. 
\begin{figure}
\begin{center}
\includegraphics[width=3in]{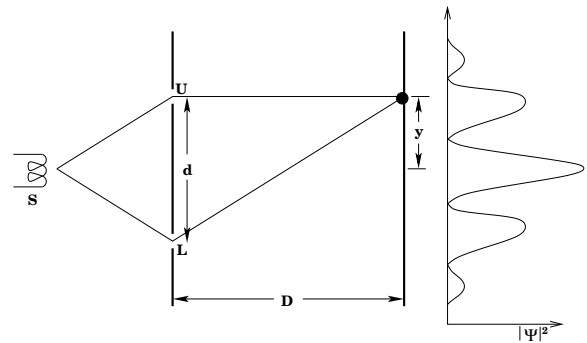}
\caption{The two-slit experiment. An electron gun at left emits an electron
traveling towards a screen with two slits, its progress in space recapitulating
its evolution in time. The electron is detected at a further screen at a position $y$ with a probability density that exhibits an interference pattern. A coarse grained set of histories for the electron is defined by specifying the slit ($U$ or $L$) through which the electron passes through and ranges $\Delta$ of the position $y$ where it is detected. In the absence of the record of  a measurement it is not possible to settle a bet on the which of these histories occurred.}   
\label{2slit}
\end{center}
\end{figure}

When there are alternatives that can be described but do not correspond to settleable bets there are two approaches to probability.   (1) Assign probabilities only to alternatives that  correspond to settleable bets in which case the usual rules of probability theory follow. (2) Assign probabilities to all alternatives, settleable or not, but allow for extensions of the usual probability theory rules for non-settleable bets. The first approach is the one usually taken in quantum theory, for example in its decoherent (or consistent) histories formulation \cite{Gri02,Omn94,Gel94}.  This paper explores the second.

The essential ideas of extended probabilities in the context of  the quantum mechanics of closed systems were given in \cite{Har04} and are easily summarized. For simplicity restrict attention to a system of particles and fields in closed box and assume that the spacetime geometry is fixed and flat. The basic input to a quantum description of the contents of the box are a Hamiltonian $H$ and an initial quantum state $|\Psi\rangle$. 
 
An exhaustive set of exclusive alternatives at one moment of time $t$ is represented\footnote{We are assuming some familiarity with the quantum mechanics of closed systems.  A pedagogical introduction to this subject in the notation used here is \cite{Har93a}.  Appendix A contains a bare bones account of decoherent histories quantum mechanics in this notation.} in the Heisenberg picture by an  exhaustive set of mutually orthogonal projection operators $\{P_\alpha(t)\}$, $\alpha= 1,2,\cdots$. An elementary example of a set of alternative {\it histories} of the closed system is  specified by a sequence of such sets $\{P^k_{\alpha_k}(t_k)\}$ at a sequence of times $t_1, t_2, \cdots, t_n$. An individual history in the set $\alpha=(\alpha_1, \cdots, \alpha_n)$ is represented by the corresponding  chain of projections
 \begin{equation}
 C_\alpha = P^n_{\alpha_n}(t_n)\cdots P^1_{\alpha_1}(t_1).
 \label{chain}
 \end{equation}
 Such chains are not generally projections themselves unless all of the members of the chain commute.\footnote{Realistic sets of histories will be {\it branch dependent} in the sense that at any one time in a given history the {\it set} of projections defining alternatives at the next time will depend on the specific previous alternatives defining the history (e.g.\cite{GH90a,GH07}). Yet more generally, sets of histories can be defined by partitions of chains like \eqref{chain} into classes whose definition is not at a discrete series of times (e.g \cite{Har91b}). To keep the exposition manageable we have ignored such generalities and restricted attention to histories of the the simple form \eqref{chain}. Most of the results extend straightforwardly to the more general cases.}
 
 Histories like ones specified by \eqref{chain} are generally {\it coarse-grained}  because alternatives are not specified at every time and because the alternatives at a given time are not projections onto a complete
set of states. {\it Fine-grained} histories consist of one-dimensional
projections at each and every time.

 We will take the extended probabilities $p(\alpha)$ for any set of histories $\{C_\alpha\}$ to be given by\footnote{This is the relation used by Goldstein and Page \cite{GP95} to define their linear positivity condition that restricts sets of histories to those for which \eqref{probs} is positive. We will rather use the relation to assign extended probabilities to all histories, without restriction. That difference should not obscure the fact that \eqref{probs} was first used a definition of probability by Goldstein and Page who also  discussed of some of its properties. Similar formulae have also been discussed in \cite{SS99,Har04,Mar05}.}
 \begin{equation}
 p(\alpha)= \Re\langle\Psi|C_\alpha|\Psi\rangle.
 \label{probs}
 \end{equation}
 As we will discuss in Section \ref{sec3},  the extended probabilities defined by \eqref{probs} satisfy all the usual probability theory rules {\it except} positivity. 
 
We will take a a set of histories to be the basis of  a {\it settleable bet} if the histories are recorded in a suitably general  sense. Specifically, a set of histories is exactly recorded if there exists a set of orthogonal projection operators $\{R_\alpha\}$ that are exactly correlated with the histories $C_\alpha$ in the state $|\Psi\rangle$. The alternative $R$'s represent the alternative values of the records. Exactly recorded histories have extended probabilities \eqref{probs}  in the range \range. The extended probabilities then  obey the usual rules of probability theory.  With a strong notion of correlation, an exactly recorded set of histories is an exactly medium decoherent set of histories and vice versa. 

Realistic sets of histories such as those defining the quasiclassical realm of everyday experience (e.g. \cite{GH07}) are not exactly medium decoherent and therefore cannot be exactly recorded. But they do satisfy these conditions to an accuracy well beyond that which a violation could be detected or, indeed, the physical situation modeled. 
  
 The relation \eqref{probs} assigns extended probabilities to {\it all} histories of a closed system. 
 These extended probabilities satisfy the usual rules of probability theory for (recorded) sets of histories that are the basis of settleable bets. But extended probabilities that are possibly negative or larger than $1$ are assigned to histories in sets that are not the basis of settleable bets. From this perspective, quantum theory consists of the specification of 
\begin{itemize}
\item An extended probability distribution on the allowed sets of fine-grained histories incorporating a theory of dynamics ($H$) and the initial quantum state ($|\Psi\rangle$).

\item A notion of coarse-graining, most generally a partition of the fine-grained histories into classes. 

\item A criterion for when a set of alternative  coarse-grained histories is the basis for a settleable bet. 

\end{itemize}

In the author's view there is no pressing need, either theoretical or experimental,  to reformulate the quantum mechanics of closed systems in terms of extended probabilities. However, such a reformulation has at least three advantages that  we develop in Sections \ref{sec5}, \ref{sec6},  and \ref{sec8}. 
 
{\it  Precision:}  The extended probabilities \eqref{probs} obey the sum rules of probabilty theory exactly.  The inevitable imprecision is rather all in the notion of settleable bet and what exactly is meant by a record. 

{\it Assuming a unique fine-grained set of histories:} It becomes possible to posit a unique set of fine grained histories together with their extended probabilities as a starting point for quantum mechanics as it is in classical theory.  Examples are particle paths in non-relativistic quantum theory, four-dimensional field configurations in the case of quantum field theory, and four-metrics and matter fields for a quantum theory of spacetime geometry.
 
{\it A starting point for generalization:}  Quantum theory leads to one way of specifying an extended probability distribution on a set of fine-grained histories of a closed system, but it may not be the only way\footnote{For an approach without fundamental  probabilities but in which the fine-grained histories manifest a non-classical logic,  see \cite{Sor06}.}. Nearby distributions that differ from that of quantum theory may be useful in parametrizing experimental tests. Quantum theory itself could be emergent from some yet more fundamental theory characterized by a different distribution. 

The natural occurrence of negative probabilities in the formalism of quantum theory has been noted by many authors and has a long, vast, varied, and interesting history. The Wigner distribution on phase space is perhaps the most familiar example \cite{Wig32,HOSW84,Zacetal06}. This and other instances of negative probability in quantum mechanics prior to 1986 were reviewed by M\"uckenheim \cite{Muc86}. Some other papers known to the author since that time are \cite{Fey87,SWS94,Khr98, Cin99,Har04,Krosum, Mar06}. The following view of Feynman \cite{Fey87}  is close in spirit to the present paper:  ``If a physical theory for calculating probabilities yields a negative probability ... we need not conclude that the theory is incorrect. ...  [A]  possibility is that the situation for which the probababilty appears to be negative is not one that can be verified.''  

The structure of the remaining parts of this paper is as follows:  Section \ref{sec2} reviews the standard argument why probabilities have to be positive for fair bets.
Section \ref{sec3} reformulates the quantum mechanics of closed systems using the extended probabilities defined by \eqref{probs} and a notion of settleable based on exactly correlated generalized records. Realistic records are discussed in Section \ref{sec7}.  Section \ref{sec4} describes the role of coarse graining plays in finding settleable bets. This formulation of quantum theory in terms of extended probabilities (EP) is compared with decoherent histories (DH) in Section \ref{sec5}.  Section \ref{sec6} formulates quantum theory in terms of a fundamental distribution for extended probabilities of a unique fine-grained set of histories. Section \ref{sec8} concludes with a discussion of the utility of extended probabilities for parametrizing experiment and extending quantum theory. 

\section{Probability}
\label{sec2}
This section briefly reviews how probabilities can be understood as instructions for betting and why they must be positive to be  bases for fair bets \cite{deF37,Cavup}. The general context is Bayesian probability theory\footnote{For a compact introduction,
see \cite{Sred05}; for complete details, see \cite{Jaynes}; for arguments for the 
necessity of the Bayesian point of view, see \cite{App04}.}.  We follow in detail the exposition in \cite{Cavup}. 

When you assert that a event $A$ has probability $p_A$ it means the following:  Suppose  a bookie\footnote{American slang for bookmaker.} offers you a bet on whether the event $A$ occurs\footnote{Or alternatively offers a bet on whether $A$ occurred,  or is occuring,  or will occur.} with a payoff $S_A$ if it does.  You will pay the bookie $p_A S_A$ and consider it a fair bet. 

The payoffs $S_A$ can be either positive or negative. Changing  their signs interchanges the roles of bookie and bettor. A negative payoff means that you pay the bookie $|S_A|$  at the end of the bet if $A$ occurs. Your payment $p_A S_A$ can also be negative meaning that the  bookie pays you $|p_A S_A|$ at the start of the bet.

Consider the simplest exhaustive set of mutually exclusive alternatives consisting of an event $A$  and the event `not $A$' denoted by $\bar A$. 
You announce to a bookie that you will accept a bet on this set of events with any payoffs $S_{A}$ and $S_{\bar A}$ that the bookie may care to set, provided that your payments for each event are $p_{A} S_{A}$ and $p_{\bar A} S_{ \bar A}$ for some numbers $p_{A}$  and $p_{\bar A}$  (probabilities) which you specify. Indeed this situation defines these probabilities. 
Your gain $G_A$ if $A$ occurs and your gain $G_{\bar A}$ if it does not are given by:
\begin{subequations}
\label{gains}
\begin{eqnarray}
G_A&=& S_A -p_A S_A -p_{\bar A} S_{\bar A},  \label{gaina} \\
G_{\bar A} &=& S_{\bar A} -p_{\bar A}S_{\bar A}- p_A  S_A  .  \label{gainb}
\end{eqnarray}
\end{subequations}

If you do not specify probabilities obeying the usual rules of probability theory the bookie will be able to chose payoffs such that you are sure to lose on every bet.\footnote{The situations where you always lose are called `Dutch books'. This is unfortunate terminology in the author's opinion but seemingly standard in English in economics and probability theory.}. Such bets are obviously not fair.  All the rules of probability theory must be satisfied for fair bets. We will illustrate the argument only to derive the requirement that probabilities must be positive for fair bet since this is the main rule we intend to modify in extending the theory. For the rest see \cite{deF37,Cavup}.

To simplify the discussion suppose the bookie decides to pay off only if $A$ occurs, ie. chooses $S_{\bar A} = 0$. Then \eqref{gains} becomes
\begin{subequations}
\label{sgains}
\begin{eqnarray}
G_A&=& (1-p_A)S_A ,  \label{sgaina} \\
G_{\bar A}&=&- p_A  S_A  .  \label{sgainb}
\end{eqnarray}
\end{subequations}
If your probability $p_A$ for $A$ is negative the bookie has only to choose $S_A$ to be negative to ensure that {\it both} $G_A$ and $G_{\bar A}$ are negative. You lose whether $A$ occurs or does not. This is not very surprising. The positive value of $p_A S_A$ means that you pay the bookie to place the bet. The negative value of  $S_A$ means that you also pay $|S_A|$ to the bookie to settle the bet if $A$ occurs. To avoid such unfair bets all your probabilities must be positive. 

Note that you lose on each bet, not on average over many.  Note also that a small negative probability does not necessarily mean that you lose a small amount. If $A$ occurs you lose $|S_A|(1+|p_A|)$ on each bet which can be as high as the bookie cares to specify. 
The bookie doesn't need to have an independent assessment of the probabilities to arrange that you always lose were you to hold negative probabilities. 
Ordinarily a bookie would make money by having a better estimate of the odds than you do. For example, if a bookie calculates probabilities with quantum mechanics while you calculate them will classical physics, the bookie will win in situations where classical physics is a poor approximation. But were you to hold negative probabilities the bookie has to know nothing else to always win. Conversely you only need to notice that all transfers are to the bookie to detect this kind of unfair bet. Indeed, you don't need the bookie at all since from your perspective the same result could be achieved  by ripping up $\$100$ bills. 

A physical theory supplies probabilities for betting on the regularities exhibited by our universe. Evidently these must be fair bets! At a minimum therefore,  the probabilities supplied by physical theory must obey the usual probability rules for bets which could be in principle be proposed and settled in the realistic universe. In particular the probabilities of records used to settle bets on which of a set of alternative histories occurs must obey these rules. If these records are exactly correlated with the histories the probabilities of the histories must obey them too. 

We next discuss extending this framework and introducing extended probabilities outside the range \range\ for non-settleable alternatives. 
These extended probabilities can be understood as part of instructions for betting. If you hold such probabilities don't bet with them! There will be no records to settle and you risk being offered an unfair wager .

\section{Extended Probabilities in Quantum Mechanics}
\label{sec3}
This section decribes how extended probabilities can be consistently assigned to every member of every exhaustive set of histories of a closed quantum mechanical system.  Then it explains how usual probabilities are recovered for sets  histories that correspond to a natural notion of settleable bets. The starting point is decoherent histories quantum theory  \cite{Gri02,Omn94,Gel94} which is very briefly reviewed in Appendix A largely to introduce the notation that is used. 

To keep the discussion manageable, we consider a closed quantum system, most generally the universe, in the approximation that gross quantum fluctuations in the geometry of spacetime can be neglected. The closed system can then be thought of as a large (say$\gtrsim$ 20,000 Mpc), perhaps expanding box of particles and fields in a 
fixed background spacetime. Everything is contained within the box, in particular galaxies, planets, observers and observed, measured subsystems, and any apparatus that measures them.  This is a model of  the most general physical context for prediction.

The fixed background spacetime means that the notions of time are fixed and that the
usual apparatus of Hilbert space, states, and operators can be employed in a quantum
description of the system.  The essential theoretical inputs to the process of prediction are the Hamiltonian $H$ governing evolution and the initial quantum state.   We assume this is  a pure state $|\Psi\rangle$.

The discussion in this section is idealized in certain respects for instance by assuming the exact decoherence of sets of histories defining settleable bets. We return to the inevitable approximations associated with more realistic situations in the next section. 

\subsection{Extended Probabilities for Histories}
Consider an exhaustive set of mutually exclusive histories $\{\alpha\}$. The two histories of an electron in the two-slit experiment  illustrated in Figure 1 are an example. A time sequence of alternative coarse-grained positions describing the various orbits that  the center of mass of the Earth might take in its progress around the Sun is another. Each history in such a set is represented by a class operator $C_\alpha$. For sets of histories defined by sets of alternatives at a series of times these will be strings of projections\footnotemark[ 2]
\ as in \eqref{chain}. As is immediate in that case [cf. \eqref{twoone}], the class operators of an exhaustive set sum to unity
\begin{equation}
\sum\nolimits_{\alpha} C_\alpha = I\ , \quad (\text{exhaustive}). 
\label{3-1}
\end{equation}

We begin with the expression [cf. \eqref{twosix}] for the probabilities $p(\alpha)$ of an {\it exactly} decoherent\footnote{Whenever `decoherent' or `decoherence' appears without defining qualification as here we mean {\it medium} decoherence as defined by \eqref{3-3}. } set of alternative histories $\{C_\alpha\}$:
\begin{equation}
p(\alpha)\equiv ||C_\alpha |\Psi\rangle||^2 = \langle\Psi| C^{\dagger}_\alpha C_\alpha |\Psi\rangle \ .
\label{deprobs}
\end{equation}
Exact decoherence means that the interference between the branch state vectors corresponding to different  coarse-grained histories vanishes [cf.\eqref{twoseven}] 
\begin{equation}
\langle  \Psi | C^\dagger_\beta C_\alpha | \Psi \rangle = 0, \quad \alpha\ne\beta . 
\label{3-3}
\end{equation} 
This is a sufficient condition for the probabilities defined by \eqref{deprobs} to obey all the usual rules of probability theory. In particular all the probabilities are manifestly positive. 

The probabilities for decoherent histories \eqref{deprobs} can be rewritten using  \eqref{3-1} and \eqref{3-3} as
\begin{equation}
p(\alpha) = \sum_{\beta} \langle\Psi | C^\dagger_\beta C_\alpha |\Psi\rangle 
\label{3-4}
\end{equation}
since the terms with $\beta\ne\alpha$ are all zero. 
Eq. \eqref{3-1}  then yields the following alternative expression for the probabilities of histories in an exactly decohering set:
\begin{equation}
p(\alpha) = \langle\Psi | C_\alpha |\Psi\rangle, \quad \text{(decohering histories)}\ . 
\label{3-5}
\end{equation}
The expression on the right hand side is neither manifestly positive nor even necessarily real. But it is real and positive for the histories of exactly decoherent sets as its derivation shows. 

We now extend the probabilities defined only for  exactly decoherent sets of alternative histories by  \eqref{3-5} to {\it all} sets of histories, decoherent or not. To do this we follow the proposal in \cite{Har04} and write\begin{equation}
p(\alpha) =\re \langle\Psi | C_\alpha |\Psi\rangle, \quad \text{(all histories)} \ . 
\label{exprobs}
\end{equation}
The real part is necessary if extended probabilitiies are to be real numbers.\footnote{Some have suggested that probabilities be extended to complex numbers e.g.  \cite{You94}.  We do not pursue this.} The relation \eqref{exprobs} reduces to \eqref{3-5} for exact decoherence since then the 
$\langle\Psi|C_\alpha|\Psi\rangle$ are real. 

The extended probabilities defined by \eqref{exprobs} are not necessarily all positive\footnote{Indeed, if the history contains more than one time there is always some $|\Psi\rangle$ for which \eqref{exprobs} is negative, e.g.  \cite{Har04}.}, nor less than 1,  but all the other requirements of usual probability theory are maintained. In particular, extended probabilities for exclusive events add and the most general form of the resulting sum rules are satisfied. A coarse graining of a set of histories $\{\alpha\}$ is a partition of the set into mutually exclusive classes $\{\bar\alpha\}$. (See the appendix for a little more detail.) Each history $\alpha$ is in some coarse grained class and  in no more than one. The operators $\{C_{\bar\alpha}\}$ representing the histories defined by the coarser grained classes are sums over operators representing the finer grained histories they contain, viz. 
\begin{equation}
C_{\bar\alpha} = \sum_{\alpha\in{\bar\alpha}} C_\alpha \  .
\label{3-7}
\end{equation} 
An immediate consequence of the linearly of \eqref{exprobs} in the $C$'s is that 
\begin{equation}
p(\bar\alpha) = \sum_{\alpha\in{\bar\alpha}} p(\alpha) \  .
\label{3-8}
\end{equation}
The most general form of the probability sum rules for exclusive alternatives is thus satisfied {\it exactly}. In particular it follow from \eqref{3-8} that
\begin{equation}
\sum\nolimits_\alpha p(\alpha) = 1 
\label{3-9}
\end{equation}
when the sum is over all histories in an exhaustive set. Allowing extended probabilities outside the range \range\  is thus the only extension of usual probabilty theory needed to assign extended probabilities to all histories that exactly obey the rest of the usual rules. 

Although it is not a part of the usual rules of probability theory, one other familiar  property besides positivity is lost in the extension from \eqref{deprobs} and \eqref{3-3}  to \eqref{exprobs}. This is the usual quantum mechanical notion of non-interacting subsystems. In quantum mechanics, a collection of $N$ unentangled non-interacting subsystems is represented by a product state
\begin{equation}
|\boldsymbol{\Psi}\rangle = |\Psi^1\rangle \otimes \cdots \otimes |\Psi^N\rangle 
\label{tenprod}
\end{equation}
in the tensor product of the Hilbert spaces of the subsystems distinguished by the superscripts.  But, as pointed out by Di\'osi \cite{Dio04}, the probabilities \eqref{exprobs} for histories $(\alpha^1, \cdots , \alpha^N)$ of the collection will not generally obey the usual product rule, that is 
\begin{equation}
p(\alpha^1, \cdots , \alpha^N) \ne p^1(\alpha^1) \cdots p^N(\alpha^N) 
\label{3-11}
\end{equation}
where the $p$ is for the collection of subsystems and the $p^i$ are computed for the individual ones, both from \eqref{exprobs}. That is because the real part of a product of numbers is not generally the product of their real parts. However, we will  show that the probabilities for {\it settleable} bets do obey this rule in the next subsection. 

\subsection{Recorded Histories}
\label{setbets}
As mentioned in the Introduction, a prescription such as \eqref{exprobs} for extending probabilities to all histories is not enough to formulate  quantum mechanics for closed systems. We must also identify those sets of alternative histories that constitute the basis for settleable bets and show that the usual rules of probability theory hold for the extended probabilities assigned to them. 

A widely accepted way of settling a bet on an event $A$ is to use a {\it record} of whether $A$ occurred that is accessible to both parties to the bet and deemed reliable by both. Indeed, bets often include a specification of what records will be used to settle them. Bets on histories of events can be settled by reliable records of the events in those histories. 

Records of a set of histories can be idealized as a set of alternatives at one time that are correlated exactly with the alternatives constituting the histories. More specifically by {\it exact records}\footnote{In previous work (eg \cite{GH95}) we have referred to operators $R_\alpha$ satisfying relations of similar character to  \eqref{records} as {\it generalized records} to emphasize that they do not necessarily have the properties of accessible records discussed below.} of a an exhaustive set of exclusive histories $\{C_\alpha\}$ we will mean a set of orthogonal projection operators $\{R_\beta(t_{\rm rec})\}$, one for each history, with $t_{\rm rec}$ greater than the last time in the history $C_\alpha$, such that 
\begin{equation}
 p(\beta,\alpha)\equiv\Re\langle\Psi | R_\beta  C_\alpha |\Psi \rangle =\delta_{\beta\alpha}\Re\langle\Psi |C_\alpha|\Psi\rangle=\delta_{\beta\alpha}p(\alpha) \ .
 \label{records}
 \end{equation}

  Sets of alternative histories for which projections $\{R_\beta\}$ satisfying \eqref{records}  {\it exist} will be called {\it exactly recorded sets of histories}. 
  Summing \eqref{records} over $\alpha$ and using \eqref{3-1} yields the following alternative expression for the extended probabilities of exactly recorded histories
\begin{equation}
p(\alpha) = \langle\Psi|R_\alpha|\Psi\rangle, \quad \text{(recorded histories)} \ . 
\label{prrec}
\end{equation}
The real part in \eqref{records} is not necessary because the expected value of any projection is always real. Furthermore, the expected value of a projection is always positive. Thus we see that the extended probabilities of exactly recorded sets of histories satisfy all the rules of usual probability theory. They are probabilities not just extended probabilities. 

The extended probabilities of recorded histories also satisfy the usual rule for the probabilities of  unentangled, non-interacting subsystems. Consider a collection of $N$ such subsystems whose state is given by \eqref{tenprod}. Records of the histories of all the subsystems ${\bf R}_{\alpha^1 \cdots \alpha^N}$ will be products of the records for the subsystems, viz 
\begin{equation}
 {\bf R}_{\alpha^1 \cdots \alpha^N}= R^1_{\alpha^1} \otimes \cdots \otimes R^N_{\alpha^N} \ .
\label{prodrecs}
\end{equation}
The extended probability $p(\alpha^1, \cdots , \alpha^N)$ will be the real part of the product of the 
$\langle\Psi^k|C^k_{\alpha^k}|\Psi^k\rangle$ for each subsystem. But since these are real for recorded histories [cf. \eqref{prrec}] we have 
\begin{equation}
p(\alpha^1, \cdots , \alpha^N) = p^1(\alpha^1) \cdots p^N(\alpha^N) 
\label{prodprobs}
\end{equation}

\subsection{Decoherent Sets of Histories and Settleable Bets}

Recorded histories for which there exist $R$'s satisfying \eqref{records} could be taken to define the notion of a settleable bet. However, a stronger condition which is satisfied approximately by realistic records is that there exist $R$'s which satisfy 
\begin{equation}
R_\alpha C_\beta |\Psi\rangle = \delta_{\alpha\beta} C_\beta |\Psi\rangle  .  
\label{strongrecs}
\end{equation}
We can call histories satisfying this condition {\it strongly recorded} and those only satisfying \eqref{records} {\it weakly recorded}. Evidently the former implies the latter. 
Realistic records are in fact strong records to a good approximation as we discuss in the next section. 

Medium decoherence  --- the vanishing of quantum interference between any pair of histories --- is a necessary condition for strong records. Specifically \eqref{strongrecs} implies [cf \eqref{twoseven}] that with $|\Psi_\alpha\rangle \equiv C_\alpha |\Psi\rangle$
\begin{equation}
\langle \Psi_\alpha | \Psi_\beta\rangle = \langle\Psi|C^\dagger_\alpha C_\beta|\Psi \rangle = 0 \quad , \quad \alpha\not=\beta \, .
\label{meddecoh2}
\end{equation}

Conversely, exact medium decoherence \eqref{meddecoh2}  implies that there exist records satisfying \eqref{strongrecs}. In fact, exact medium decoherence implies that there are many  sets $\{R(t_{\rm rec})\}$ of exhaustive and exclusive projections satisfying \eqref{strongrecs}.
 The simplest case is to take the Heisenberg picture projections on the subspaces spanned by the $\{|\Psi_\alpha\rangle\}$, $R_\alpha \equiv {\rm Proj}\{|\Psi_\alpha\rangle\}$,  and assign them a time 
$t_{\rm rec}$ later than the last time in the history $C_\alpha$. The orthogonality of the $|\Psi_\alpha\rangle$ implied by medum decoherence ensures that these $R_\alpha$ are orthogonal. 

Requiring exact strong records and exact medium decoherence are thus the same thing.
The {\it existence} of strong records reduces to a calculable condition \eqref{meddecoh2} on the sets of alternatives alone. That is a considerable advantage over a condition formulated just in terms of the existence of records like \eqref{records}.

{\it In this paper,  we will take the existence of strong records or equivalently  medium decoherence as the minimal requirement for  settleable bets.} 
When account is taken of necessary approximations (Section III.D), the notions of extended probability supplied by \eqref{exprobs} and medium decoherence defined by \eqref{meddecoh2} can be taken as the fundamental elements in a quantum mechanics of closed systems. 

\section{Approximate Decoherence and Realistic Records}
\label{sec7} \label{approx}

\subsection{Approximate Decoherence}

Perhaps the most important examples of sets of alternative coarse-grained histories are those which describe the wide ranges of place, epoch, and scale on which the laws of classical physics hold approximately in this quantum universe (e.g. \cite{Har94b,GH07}.)  The sets of histories defining this quasiclassical realm  do not decohere exactly. Rather they decohere to an approximation well beyond the accuracy to which we could conceivably check their predictions,  or, indeed, beyond the accuracy the physical situation in which they apply can be manageably modeled. Crude estimates of the overlap $\epsilon$  between branches of this quasiclassical realm \cite{Har91a,Harup} give numbers of the {\it qualitative} form $\epsilon\sim 10^{-E}$ where $E\sim 10^{20}$.

Approximately decohering sets of histories cannot be exactly recorded. But we can expect records $\{R_\alpha\}$ that are correlated with histories to an accuracy comparable to $\epsilon$, that is
\begin{equation}
|\langle\Psi|R_\alpha|\Psi\rangle -\Re\langle\Psi|C_\alpha|\Psi\rangle|< \epsilon. 
\label{diff}
\end{equation}
This means that the extended probabilities for histories could be outside the range \range, but not by much. For example we could have negative values of   $\Re\langle\Psi|C_\alpha|\Psi\rangle$  only if 
\begin{equation}
|\langle\Psi|R_\alpha|\Psi\rangle|< \epsilon, \quad |\Re\langle\Psi|C_\alpha|\Psi\rangle|< \epsilon. 
\label{smallp}
\end{equation}
For all practical purposes these deviations from the range \range\ are negligible. In fact, as discussed in Section \ref{sec4}, a tiny amount of coarse graining can be expected to give sets of histories with probabilities satisfying all the usual rules.

Some degree of  approximation seems to be necessary to formulate a quantum mechanics of alternative histories of a closed system. Perhaps this is not surprising for a theory whose outputs are probabilities. The application of probability theory inevitably involves subjective choices. For instance, how small does a probability for an event have to be before we act as if it were precluded? This subjective uncertainty permits the theoretical imprecision we have discussed. In the present instance,
extended probabilities obey all the rules of probability theory exactly except for being in the range \range~. The approximation comes in providing a realistic notion of settleable bet. Intuitively that seems a natural place to put it.

\subsection{Realistic Records}
\label{realrecs}

The identification of recorded sets of histories with settleable bets did not require that the records were accessible to human IGUSes, only that they exist in the sense of \eqref{strongrecs}. 
In many instances the records that arise from mechanisms of  decoherence are not accessible to us. In the classic example of Joos and Zeh \cite{JZ85},  a dust grain is in a superposition of positions deep in interstellar space. Alternative coarse-grained histories of its position are decohered though its interaction with CMB photons scattering off it. 
The states of the  scattered photons contain a record of the histories of grain's position. But this record  is not accessible to us. The photons left with the speed of light at the time of scattering. 

The histories of the quasiclassical realm are defined by averages of conserved quantities such as energy, momentum, and number over suitable spatial volumes. A record of the history may be contained in degrees of freedom internal to these volumes or in the variables themselves \cite{Hal99, BH99} but we typically can't get at them.

For these reasons, human bettors will usually have requirements beyond \eqref{strongrecs}  on the records used to settle their bets. For example, human IGUSes may want to restrict the notion of settleable bet to sets of histories whose records have any or all of the following properties: (1) The records are in variables other than those constrained by the coarse grainings defining the histories.  (2) The records are in quasiclassical variables can be immediately registered by our senses. (3) The records persist for a time longer than that required for human perception ($\sim .1$s). (4)  The records are of the outcomes of measurements. (5) The records are accessible, right here, right now on Earth with current technology at bearable cost.
Such records will typically be less precisely correlated with the histories than is in principle possible consistent with limitations like \eqref{diff}. 

While individual bettors may want restrict the notion of record by criteria beyond \eqref{strongrecs},  there is no reason to incorporate such restrictions in a fundamental formulation of quantum theory. Indeed, there is every reason not to\footnote{Strong decoherence \cite{GH95} may yield a more realistic but still general notion of record.}. All the notions (1)-(5) are human specific and none of them admits of a precise formulation.  Requiring strong records in the general sense of  \eqref{strongrecs} is a precise criterion for  identifying sets whose extended probabilities obey the usual rules of probabiltiy theory and that therefore can be  the basis of settleable bets.

\section{Coarse Graining and Settleable Bets}
\label{sec4}
Even a single negative extended probability in a set of histories enough to show that its alternatives are not the basis for a settleable bet. But not every set of histories with positive probabilities is the basis of a settleable bet. The set must decohere for that, and there are sets of histories with all positive extended probabilities that do not decohere. Indeed, these  are just the linearly positive sets \cite{GP95} that do not decohere. A number of simple examples were given in \cite{Har04}. 

\subsection{Coarse Graining Leads to Decoherence}
Coarse graining is a route to decoherence and therefore to settleable bets. We can see this using a simple measure of how far away a set of histories is from decohering. For  this it is convenient to introduce the {\it decoherence functional} of a set of alternative coarse-grained histories. When a pure initial state is assumed, as here, this is
\begin{equation}
D(\alpha,\beta) \equiv \langle \Psi_\alpha|\Psi_\beta\rangle = \langle\Psi|C^\dagger_\alpha C_\beta |\Psi\rangle .
\label{decohfnal}
\end{equation}
A set decoheres exactly when the off-diagonal elements of $D$ vanish. A convenient measure of the absence of decoherence is then
\begin{equation}
\dec(D) \equiv \sum_{\alpha \ne \beta} |D(\alpha,\beta)| .
\label{decmeas}
\end{equation}
This vanishes when the set of histories is exactly decoherent. 

Coarse graining decreases $\dec(D)$. Let $\{\alpha\}$ be a set of histories and $\{\bar\alpha\}$ coarse graining of it, that is, a partition of $\{\alpha\}$ into an set of mutually exclusive classes. (See the Appendix for further discussion and notation.) Then
\begin{equation}
\bar D(\bar\alpha,\bar\beta) = \sum_{\alpha \in \bar\alpha} \sum_{\beta \in \bar\beta}
 D(\alpha,\beta) .
 \label{cgD}
 \end{equation}
 Evidently 
 \begin{equation}
 \dec(\bar D) \le \dec(D)
 \label{smallerdec}
 \end{equation}
 Continued coarsed graining will drive $\dec(D)$ towards zero, decoherence,  and settleable bets. Further coarse-grainngs of decoherent sets remain decoherent. We  will  illustrate this with two simple models, but first we discuss the effect of coarse graining on extended probabilities. 

\subsection{Coarse Graining and Extended Probabilities} 
Coarse graining doesn't always drive sets of histories with extended probabilities outside the range \range \ to ones with probabilities within that range.  For example consider a set with extended probabilities $p(\alpha)$. Suppose that some of these are less than zero and some are greater than $1$, all summing to unity. Coarse grain this set by partitioning it into two classes. In the first put all the histories with negative extended probabilities. In the second put all the rest. One history has an extended negative probability, the other has one larger than $1$. This set has no  extended probabilities even in the range \range. The only further possible coarse-graining is to combine both sets into the trivial set with one history with probability $1$. 

More interesting results are obtained with restricting to more sensible coarse-grainings.
For instance, restrict attention to sets of histories defined by chains of sets of projections as in \eqref{chain} that describe sequences of events at a series of times. Restrict attention to coarse-grainings that preserve this property. The extended probability of a chain is not necessarily positive but it is always less than $1$
\begin{equation}
p(\alpha) = \Re[ \langle \Psi | P^n_{\alpha_n}(t_n)\cdots P^1_{\alpha_1}(t_1)|\Psi\rangle]  \le \langle\Psi|\Psi\rangle =1. 
\label{probchains}
\end{equation}
In this case, coarse-graining will always decrease the total negative extended probability
\begin{equation}
p_{\rm tot. neg.} \equiv \sum_{p < 0} p(\alpha) .
\label{totneg}
\end{equation}
That is  because any coarser grained set that includes a negative extended probability history along with some positive probabilities will make a smaller contribution to the negative side of the ledger than those histories did before.

\subsection{Illustrative Examples}

\subsubsection{The Two-Slit Experiment}
\begin{figure}  
\includegraphics[width=3in]{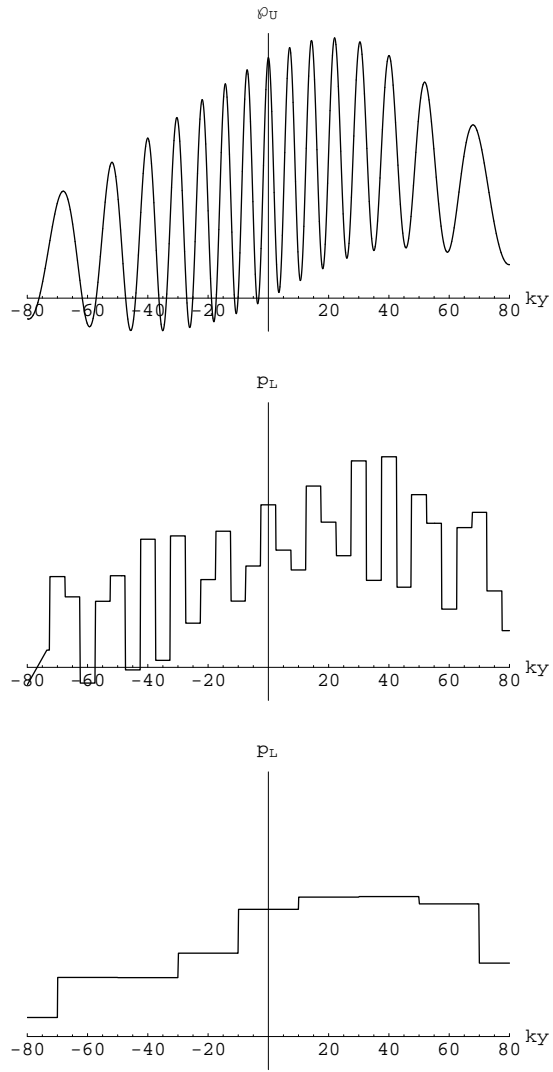}
\caption{Coarse-graining in the two-slit experiment shown in Figure \ref{2slit}. The top figure shows the extended probabiity density $\wp(y,U)$ for the electron to pass though the upper slit $U$ and arrive at a point $y$ on the screen for the values $kD=kd=60$ where $k$ is the wave number of the electron. The amplitude $a$ has been assigned arbitrarily  which is why there is no scale on the vertical axis. The extended probability density is negative for some range of $y$. The next two figures show the effect of coarse graining by dividing $y$ into bins of equal width $\Delta$. In the first figure $k\Delta = 5$ and in the second it is $20$. The final coarse graining  has all positive  extended probabilities  even though there is still significant interference between the histories. }
\label{2slitprobs}
\end{figure}

We consider the classic two-slit experiment illustrated in Figure \ref{2slit}.
We make the usual idealizations, in particular assuming that
the electrons are initially in narrow wave packets moving
in the horizontal direction $x$, so that their
progress in $x$ recapitulates their evolution in time.  We assume that the
source is far enough from the first screen that these wave packets can be
analyzed into plane waves propagating in the $x$ direction with a
distribution peaked about a wave number $k$. Then, to a good approximation, we
can calculate the amplitudes for detection by considering just the plane
wave with peak wave number $k$. A sketch of that analysis follows; for more detail see \cite{Har04}. 

Consider the coarse grained history $(i,U)$ in which the electron passes through the upper slit $U$ and arrives in one of an exhaustive set of exclusive ranges $\{\Delta_i\}$ of $y$.  From \eqref{exprobs}, the extended probability for this history can be written
\begin{equation}
p(i,U)=\Re ( \langle \Psi|\Psi_{(i,U)}\rangle) 
\label{2slit-1}
\end{equation}
where $|\Psi_{(i,U)}\rangle = C_{(i.U)}|\Psi\rangle$ is the branch state vector for this history. Similar expressions hold for  $p(i,L)$, $|\Psi_{(i,L)}\rangle$, etc.  Introducing configuration space representatives this can be written
\begin{subequations}
\begin{align}
p(i,U) = &\int_{\Delta_i} dy \Re[\Psi^*(y)\Psi_U(y)] \label{2slit-1a} \\
                       \equiv     & \int_{\Delta_i} dy \wp(y,U) 
\label{2slit-2b}
\end{align} 
\end{subequations}
where  $\Psi(y) = \Psi_U (y) +\Psi_L (y)$ and  $\wp(y,U)$ is  the extended probability density for passing through the upper slit and arriving at point $y$ on the screen. 

The amplitude $\Psi_U(y)$ is approximately given by
\begin{equation}
\Psi_U(y) \equiv ae^{ikS_U(y)}/S_U(y)
\label{fourone}
\end{equation}
where
\begin{equation}
S_U(y)\equiv \left[\left(d/2 -y\right)^2 + D^2\right]^{\frac{1}{2}}
\label{fourtwo}
\end{equation}
is the distance from the upper slit to the point on the detecting screen
labeled by $y$ and $a$ is a constant amplitude. 
With this, the extended probability {\it density} $\wp(y, U)$  to arrive at
$y$  having passed through $U$ defined in \eqref{2slit-2b} works out to be
\begin{equation}
\wp (y, U) = \frac{|a|^2}{S_U}\ \left\{\frac{1}{S_U} + \frac{1}{S_L}
\cos [k(S_L-S_U)]\right\}.
\label{foursix}
\end{equation}

To see the effect of coarse graining on the extended probabilities $p(i,U)$ consider the simple case shown in Figure \ref{2slitprobs}  when all the ranges have the same size $\Delta$. The top figure shows $\wp(y,U)$ which has a range of negative values showing that the histories $(y,U)$ and $(y,L)$ are not the basis of a settleable bet. 
With increasing coarse graining --- increasing $\Delta$ --- the extended probabilities eventually become positive even though there is significant  quantum interference ($\sim D/k\Delta d$) between the two histories. Continued coarse graining  will  reduce that interference until the medium decoherence condition is satisfied
\begin{equation}
\int_\Delta dy\, {\rm Re} \left[\Psi^*_L (y) \Psi_U (y)\right] \approx 0.
\label{fourten}
\end{equation}
That requires a $\Delta$ much larger than the spacing between the interference
fringes and much larger than that required merely for positive probabilities, but then the histories are the basis of a settleable bet.

\subsubsection{The Three Box Example}
The three-box example introduced by Aharonov and Vaidman \cite{AV91} for
other purposes provides another example of effect of coarse-graining on extended  extended probabilities. 

Consider a particle that can be in one of three boxes, $A$, $B$, $C$ in
corresponding orthogonal states $|A\rangle$, $|B\rangle$, and $|C\rangle$.
For simplicity, take the Hamiltonian to be zero, and suppose the system to
initially be in the state
\begin{equation}
|\Psi \rangle \equiv \frac{1}{\sqrt{3}}\ (|A\rangle + |B\rangle +
|C\rangle).
\label{foursixteen}  
\end{equation}
Consider further a state $|\Phi\rangle$ defined by 
\begin{equation}
|\Phi\rangle \equiv \frac{1}{\sqrt{3}}\ (|A\rangle + |B\rangle -
|C\rangle)
\label{fourseventeen}
\end{equation}
and denote the projection operators on $|\Phi\rangle$, $|A\rangle$,
$|B\rangle$, $|C\rangle$ by $P_\Phi$, $P_A$, $P_B$, $P_C$ respectively.
Denote with $\bar A$ the negation of $A$ (``not in box $A$'') represented
by the projection $P_{\bar A} = I-P_A$. The negations $\bar \Phi, \bar B,
\bar C$ and their projections $P_{\bar \Phi}, P_{\bar B}$, and $P_{\bar C}$
are similarly defined. 

As a simple example consider a set of histories defined by whether the particle is in box $A$, $B$, or $C$ at a certain time {\it given} that it is in the state $|\Phi\rangle$ at a later time. Since $H=0$ the values of these times are not important, only their order. This set can be thought of as  a simplified version of a {\it three}-slit experiment analogous to the two-slit one considered above (see, e.g. \cite{Sorsum}). 

There are three histories in this set distinguished by which box the particle is in at the intermediate time. Their class operators are 
\begin{equation}
P_\Phi P_A, \quad P_\Phi P_B, \quad P_\Phi P_C.
\label{3box-1}
\end{equation}
Their extended probabilities are given by \eqref{exprobs}, e.g. 
\begin{equation}
p(\Phi,A) = \Re \langle \Psi| P_\Phi P_A|\Psi\rangle . 
\label{3box-2}
\end{equation}
A little calculation gives 
\begin{equation}
p(\Phi,A)= 1/9, \quad p(\Phi,B)= 1/9, \quad p(\Phi,C)= -1/9.
\label{3box-3}
\end{equation}
The conditional extended probabilites for which box the particle was in  given $|\Phi\rangle$,  e.g $p(A|\Phi) \equiv p(\Phi,A)/p(\Phi)$,  are
\begin{equation}
p(A|\Phi)= 1, \quad p(B|\Phi)= 1, \quad p(C|\Phi)= -1.
\label{3box-4}
\end{equation}
Evidently, there is no settleable bet possible on which of the three boxes the particle was in at the intermediate time given $|\Phi\rangle$. Thus, there is no contradiction between the unit probabilities assigned to the exclusive alternatives $A$ or $B$ in \eqref{3box-4} \cite{GrH97}. 

This set of histories can be coarse-grained in three different ways by combining two of the three histories into one coarser grained one. For instance, the set of histories defined by whether the particle is or is not in box $A$ at the intermediate time given $|\Phi\rangle$ at the later time is represented by the chains $P_\Phi P_A$ and
 $P_\Phi P_{\bar A}$.  These three mutually incompatible sets have the extended probabilities
\begin{subequations}
\begin{eqnarray}
&p(A|\Phi)=1, \quad  &p(\bar A|\Phi) = 0,   \label{3box-5a} \\
&p(B|\Phi)=1, \quad &p(\bar B|\Phi) = 0,  \label{3box-5b} \\
&\  \ p(C|\Phi)=-1, \quad &p(\bar C|\Phi) = 2.  \label{3box-5c}
\end{eqnarray}
\end{subequations}

It is easy to check that the first two coarse grained sets are medium decoherent. Thus it  is possible to make a settleable bet on the coarse grained alternatives of  whether the particle was in box $A$ or not in $A$ at the intermediate time, in $B$ or not in $B$, but not on $C$ or not in $C$.  Needless to say in this highly simplified model  the generalized records associated with the settleable alternatives are trivial and nothing like those that might be available in a realistic three-slit experiment.

\section{Comparison with Decoherent Histories Quantum Theory}  
\label{sec5} 
This section compares the formulation of quantum mechanics in terms of extended probabilities (EP) with the decoherent histories quantum mechanics (DH) summarized in the Appendix. DH and EP agree on the predictions for exactly decoherent settleable bets. The probabilities are the same because of the equality between \eqref{3-5} and \eqref{exprobs}. 
\EP can be viewed as extension of \DH that assigns extended probabilities to all sets of histories. However, \DH and \EP differ on how they incorporate the quantum mechanical arrow of time and how they deal with approximate decoherence. This section describes these differences. There are no experimental  differences, but they may be different starting points for extending or modifying quantum theory as we discuss in Section \ref{sec8}. 

\subsection{Arrows of Time}
In \DH the expression for the probabilities of time histories is [cf \eqref{twosix}]
\begin{equation}
p(\alpha) = \parallel |\Psi_\alpha\rangle\parallel^2 = 
\parallel C_\alpha |\Psi\rangle\parallel^2\, .
\label{twosix1}
\end{equation}
This is not not time neutral. 
The state $|\Psi\rangle$ is on one end of the chain of projections in $C_\alpha$ [cf. \eqref{chain}]  and there is nothing on the other end. This time asymmetry is called the quantum mechanical arrow of time. 
The end of the chain with the state is conventionally called {\it the past} and the other end is called {\it the future.} Histories are time ordered with the earliest times assigned to the past\footnote{This order can be reversed by a $CPT$ transformation since field theory is invariant under $CPT$, but that does not alter the fact that there is an asymmetry between the past and the future.}~\footnote{Quantum mechanics  can be formulated time neutrally  using both initial and final conditions \cite{ABL64,GH93b}. Then both the quantum mechanical and thermodynamic arrows of time arise from the difference between a special initial condition $|\Psi\rangle$ and a final condition of indifference represented by a density matrix proportional to the unit matrix. }. 

By contrast, as pointed out by Goldstein and Page \cite{GP95}, the extended probability formula for \EP \eqref{exprobs} is time neutral. The extended probabilities $p(\alpha)$ are the same as whether the chains $C_\alpha$ have one time order or the reverse $C^\dagger_\alpha$. 
Rather time asymmetry in \EP arises from the definition of a settleable bet as a recorded set of histories. In expression \eqref{records}, the record operators are after the histories not before. 

This time asymmetry is consistent with the second law of thermodynamics. Physical records in the quasiclassical  realm are often created by irrreversible processes in which entropy increases. An impact crater on the moon, an ancient fission track in mica, and the printed ink of on this page are all examples. Consistent with the second law, records of events in histories are more likely to be at the end of history furtherest from an initial condition of low entropy than at the beginning.\footnote{For a slightly amplified discussion of this see \cite{Har04} and also \cite{GH07}.}. 

In \DH there is an intrinsic quantum mechanical arrow of time which is  consistent with the second law if the entropy was low in the early universe. In \EP the quantum mechanical arrow of time arises from the second law in a fundamental framework which is time neutral. 

\subsection{Approximate Decoherence} 
We next turn to the comparison between \DH and \EP when decoherence is not exact. 
For realistic sets of alternatives such as those defining the quasiclassical realm (e.g.\cite{Har94b,GH07})  the decoherence condition \eqref{meddecoh2} will only be  satisfied to an excellent approximation
\begin{equation}
\langle\Psi_\alpha|\Psi_\beta\rangle\approx 0 \quad , \quad \alpha\not=\beta \, .
\label{approxdecoh}
\end{equation}
When decoherence is approximate the probabilities for histories defined by \EP and \DH will not coincide exactly. Rather
\begin{equation}
p^{\rm DH}(\alpha)= p^{\rm EP}(\alpha) - \Re \sum_{\beta \ne\alpha} \langle \Psi| C^\dagger_\beta C_\alpha |\Psi\rangle .
\label{probsdhep}
\end{equation}
While $p^{\rm EP}(\alpha)$ obey the probability sum rules \eqref{3-8} the \DH probabilities will obey them only to an approximation defined by the standard of decoherence. 

For realistic sets of histories such as those defining a quasiclassical realm we expect to have the probabilities of \DH and \EP agree, and the rules of probability theory satisfied to an accuracy fare beyond that to which the probabilities can be checked, the sum rules verified, or, indeed, the physical situation correctly modeled (e.g.\cite{Har91a}, Sec. II.11). Put differently we expect that the loss from the unfair bets that are permitted because the probability sum rules are slightly violated to be negligable for any set of bets that could be settled in the accessible universe \cite{Harup}. Thus for all practical purposes \DH and \EP are equivalent. 

However,  \DH and \EP manage the approximations that are inevitable in any formulation of quantum theory differently. In \DH the sum rules of probability theory are approximately satisfied as are the correlations defining records.  In \EP the sum rules of probability theory are satisfied exactly although at the expense of introducing extended probabilities with values outside the range $0$ to $1$. The approximations are confined to the level of correlation required between histories and the records by which bets are settled. Some may find this a more natural and satisfactory way of organizing inevitable  theoretical uncertainty\footnote{We can also compere EP with linear postivity LP. Goldstein and Page \cite{GP95}  used \eqref{exprobs} to introduce a very weak notion of decoherence called {\it linear positivity}. Eq. \eqref{deprobs} for probabilities was replaced by \eqref{exprobs} and 
the decoherence condition \eqref{3-3} was replaced by the linearly positive condition 
\begin{equation*}
\re \langle\Psi | C_\alpha |\Psi\rangle \ge 0,   \quad \text{all}\ \ \alpha \ . 
\end{equation*}
for {\it all} histories in the the set $\{\alpha\}$. Evidently from \eqref{3-5} medium decoherence implies linear positivity but not the other way around. Probability assignments in quantum theory would be restricted to linearly positive sets of histories. We have taken medium decoherence as the minimal requirement for a settleable bet. LP extends probabilities beyond that \cite{Har04} to sets of histories with positive extended probabilities are not the basis of settleable bets. Nothing seems to be lost and a some theoretical unity gained by extending the notion further to all histories as here.}.

\section{The Fundamental Distribution}
\label{sec6}
This  paper has introduced a new understanding of extended probabilities lying outside the range of $0$ to $1$. They represent alternatives that are not the basis for settleable bets because there are no records to settle them. With this understanding, quantum mechanics can be formulated as a stochastic theory predicting extended probabilities by positing two fundamental ingredients:
\begin{itemize}

\item A unique set of alternative fine-grained histories $\{h\}$. 

\item A fundamental distribution $w(h)$ giving the extended probability for each fine-grained history in the set. 

\end{itemize} 
A set  of coarse-grained histories is a partition of the set of $h$'s into an exhaustive set of exclusive classes $c_\alpha$, $\alpha = 1,2,\cdots$. The extended probability of each coarse-grained history $c_\alpha$ is 
\begin{equation}
p(\alpha) = \sum_{h\in c_\alpha} w(h) . 
\label{6-1}
\end{equation}

\EP quantum mechanics is defined by fundamental distributions  of the form \eqref{exprobs} incorporating both state and dynamics. To give a concrete example suppose that the $h$'s were four dimensional field configurations $\phi(\vec x,t)$ on a fixed, background spacetime. The fundamental distribution on field configurations between an initial time $t'$ and a final time $t''$ would be 
\begin{align}
w[&\phi(\vec x,t)] =  \nonumber \\
&\re\{ \Psi^{*}[\phi''(\vec x), t'') \exp\{i S[\phi(\vec x,t)]/\hbar\}
\Psi[\phi'(\vec  x),t')\}
\label{6-2}
\end{align}
where $\phi(\vec x, t') = \phi'(\vec x)$ and $\phi(\vec x, t'') = \phi''(\vec x)$. The functional $S[\phi(\vec x,t)]$ is the action for field configurations between $t'$ and $t''$. The Schr\"odinger picture  wave functional  $\Psi[\phi'(\vec x), t')$ is the state  in the field representation at $t'$ and similarly for the functional at $t''$.  The usual measure for functional integration over fields is prescribed to define the sums in \eqref{6-1}. The oscillation of the exponential in \eqref{6-2} shows that $w[\phi(\vec x, t)]$ will be negative for some fine-grained histories so that no bet on fine-grained histories is possible. This is the fundamental distribution defining usual quantum field theory.

The assumption of a unique set of fine-grained histories prefers one representation over others in formulating quantum theory --- the field representation in the case of \eqref{6-2}. Transformation theory relating this preferred representation  to others  would be a derived or approximate notion. This is an old idea already implicit in Feynman's sum-over-histories formulation of quantum mechanics (e.g. \cite{Fey48,FH65}) for which the fundamental distribution \eqref{6-2} is a particular expression.\footnote{The idea of a preferred representation is developed for non-relativistic quantum mechanics in \cite{FH65}. It has advocated by many in various situations. See, e.g. \cite{Sorsum,Har93c}.}  Similarly, all the usual apparatus of quantum theory especially its 3+1 formulation in terms of the unitary evolution of states on spacelike surfaces is not posited but rather derived from this fundamental distribution \eqref{6-2} \cite{FH65}.\footnote{Such a derivation requires a fixed background spacetime foliated by a  spacelike surfaces on which the histories $\{h\}$ are single valued. Neither of these prerequisites  can be expected when spacetime geometry is a quantum dynamical variable (e.g. \cite{Har07a,Har95c}). In such cases the Feynman's spacetime formulation of quantum theory is more general than a 3+1 formulation.}

In classical physics probabilities represent ignorance of the true physical situation. We use these probabilities  to bet on what this situation is, and settle bets by determining it. 
The characteristic feature of quantum mechanics is that there are alternative situations that can be described but for which it is not possible to settle a bet on which is the true situation. Extended probabilities extend the classical notion of ignorance to include all such non-settleable bets.\footnote{A fundamental distribution like $w[\phi(\vec x, t)]$ in \eqref{6-2} is like a classical distribution but  valued in extended probabilities. 
The variables $\phi(\vec x, t)$ are then like local hidden variables. The limitations of Bell's inqualities for local hidden variable theories do not hold if the probabilities are not positive (e.g \cite{HHK96,RS01}).  Can we then say that one history is `realized' and the extended probabilities represent our ignorance of what it is? The answer depends on what is meant by `realized'. If it required that there be a way of checking which history is realized to make this assertion then the answer is `no'. The extended probabilities mean that there is no way of doing this. But if only consistency with our experience is required to make the assertion then the answer is `yes'. For the author's views on this kind of question see  \cite{Har07b}.}

\section{Testing, Extending and Modifying Quantum Theory}
\label{sec8}
Why bother formulating the quantum mechanics of closed systems in different ways? The answer is not in order to find different predictions for experiment. By definition these are the same as they are for \EP and \DH as discussed in Section \ref{sec5}. Were they different we would have different theories that could be distinguished experimentally.\footnote{ Indeed, the more restricted approximate quantum mechanics of measured subsystems (the Copenhagen formulation) is adequate for the prediction of all laboratory measurements. Neither \EP or \DH is necessary for that.} 

One reason for new formulations is that they may be conceptually clearer to individual scientists because they are closer to incorporating their theoretical prejudices. \EP for instance is closer to classical physics in many ways than some more standard formulations as we have discussed. 

But a deeper and more important  reason is that reformulations of quantum mechanics provide different starting points for extending, modifying, and generalizing it. Experiment provides one motivation for this, and the search for a final theory another.

\subsection{Experimental Tests of Quantum Mechanics}

Today there is not one shred of experimental evidence against quantum mechanics and much to be found for it on scales ranging roughly from those probed by the highest energy particle accelarators ($10^{-17} \cm$) to delicate experiments on condensed matter systems ($10^{-5} \cm$). That is a wide range of scales but still small compared to the range of $10^{-33} \cm$ to $10^{28}\cm$ that characterize the phenomena considered  in contemporary physics. Recent experiments have extended the range on which quantum mechanics has been or will be  tested (e.g.  \cite{Moo00, Zei02, Moo03, Fri03,Mar03}). To motivate and analyze future experiments that probe quantum mechanics at new scales it would be very useful to have alternative theories.  These should agree with quantum mechanics where it has been tested so far,  but differ from it on scales where it has not yet been tested and be consistent with the rest of modern physics such as special relativity.  But as Steve Weinberg puts it \cite{Wei92}: ``It is  striking that so far it has not been possible to find a logically consistent theory that is close to quantum mechanics other than quantum mechanics itself''. 

\EP may help. It reduces quantum theory to the specification of a certain fundamental distribution of  $w(h)$ of extended probabilities (e.g \eqref{6-2}). It should be possible to construct alternative theories by considering distributions that are close to those of quantum theory but differ from on the length scales where it has not been tested. 

\subsection{Emergent Quantum Theory}

Generalizations of  quantum theory are necessary for a unified final theory of our particular universe and of all the fundamental dynamical interactions that occur within it. 
The textbook quantum mechanics of measurements and observers already has to be generalized to apply to a closed system like the universe. \DH and \EP provide such generalizations. But quantum theory has to be generalized further to accommodate quantum spacetime when the fixed spacetime geometries in which unitary state evolution operates are no longer available. Generalizations have been sketched \cite{Har95c,Har07a} but the generalizing the fundamental distribution of \EP may provide others. 

The structure of quantum mechanics incorporates notions of spacetime \cite{Har07a} and is closely linked to spacetime ideas such as the notions of initial and final conditions. Could quantum mechanics itself emerge from something deeper along with an emergent notion of spacetime? The generalizations permitted by extending the fundamental distribution provided by \EP may help provide an answer.

\acknowledgments
Discussions with Carl Caves, Fay Dowker,  Murray Gell-Mann, Adrian Kent,  Simon Saunders, Ruediger Schack, Rafael Sorkin, Mark Srednicki, and David Wallace were  useful. This work was supported in part by the National Science Foundation under grant PHY05-55669.

\appendix

\renewcommand{\theequation}{\Alph{section}.\arabic{equation}}

\section{The Quantum Mechanics of Closed Systems}

Largely to explain the notation this appendix  gives a bare-bones account of some essential elements of the modern synthesis of ideas constituting  the decoherent histories quantum mechanics of closed systems \cite{Gri02, Omn94, Gel94}. We fix attention on themodel universe in a box described at the beginning of \ref{sec2}. 
 
The most general objective of quantum theory is the prediction of the probabilities of
individual members of sets of coarse-grained alternative histories of the closed
system. For instance, we might be interested in alternative histories of the center-of-mass
of the Earth in its progress around the Sun, or in histories of the correlation between the
registrations of  a measuring apparatus and a  property of the subsystem. 

Alternatives at one moment of
time can always be reduced to a set of yes/no questions.  For example, alternative positions 
of the Earth's center-of-mass can be reduced to asking, ``Is it in this region -- yes or no?'',
``Is it in that region -- yes or no?'', etc. An exhaustive set of yes/no alternatives is
represented in the Heisenberg picture by an exhaustive set of orthogonal projection operators 
$\{P_\alpha(t)\}$,
$\alpha = 1, 2, 3 \cdots$.  These satisfy
\begin{equation}
\sum\nolimits_\alpha P_\alpha(t) = I, \  {\rm and}\ P_\alpha(t)\, P_\beta (t) =
\delta_{\alpha\beta} P_\alpha (t) \ ,
\label{twoone}
\end{equation}
showing that they represent an exhaustive set of exclusive alternatives.  In the Heisenberg
picture, the operators $P_\alpha(t)$ evolve with time according to 
\begin{equation}
P_\alpha(t) = e^{+iHt/\hbar} P_\alpha(0)\, e^{-iHt/\hbar}\, .
\label{twotwo}
\end{equation}
The state $|\Psi\rangle$ is unchanging in time.

An important kind of set of histories\footnotemark[2] is specified by a series of sets of single time
alternatives $\{P^1_{\alpha_1} (t_1)\}$,
$\{P^2_{\alpha_2} (t_2)\}, \cdots$, $\{P^n_{\alpha_n} (t_n)\}$ at a sequence of times
$t_1<t_2<\cdots < t_n$.  The sets at distinct times can differ and are distinguished by the
superscript on the $P$'s. For instance, projections on ranges of position might be followed
by projections on ranges of momentum, etc.  An individual history $\alpha$ in such a set is a
particular sequence of alternatives $(\alpha_1, \alpha_2, \cdots, \alpha_n)\equiv \alpha$
and is represented by the corresponding chain of projections called a {\it chain or
class operator}
\begin{equation}
C_\alpha\equiv P^n_{\alpha_n} (t_n) \cdots P_{\alpha_1} (t_1)\, .
\label{twothree}
\end{equation}

A set of histories like one specified by \eqref{twothree} is generally {\it coarse-grained}  because alternatives are specified
at some times and not at every time and because the alternatives at a given time are typically 
projections on subspaces with dimension greater than one and not projections onto a complete
set of states. Perfectly  {\it fine-grained} sets of histories consist of one-dimensional
projections at each and every time.

Operations of fine and coarse graining may be defined  on sets of histories.  A set of histories $\{\alpha\}$ may
be {\it fine -grained} by dividing up each class into
 an exhaustive set of exclusive subclasses $\{ \alpha' \}$.  Each subclass  consists of some number of histories in a coarser-grained class, and
every finer-grained subclass is in some  class.   {\it Coarse graining} is the operation of uniting subclasses of histories into bigger classes.  Suppose, for example, that the position
of the Earth's center-of-mass is specified by dividing space into cubical regions of a
certain size. A coarser-grained description of position could consist of
larger regions made up of unions of the smaller ones. Consider a set of histories with class operators $\{C_\alpha \}$ and a coarse graining with class operators
 $\{ {\bar C}_{\bar\alpha} \}$  . The operators $\{{\bar C}_{\bar\alpha}\}$  are then related to the  operators $\{C_\alpha\}$ by summation, {\it viz.}
\begin{equation}
\bar C_{\bar\alpha} = \sum_{\alpha\in\bar\alpha} C_\alpha \ ,
\label{twofour}
\end{equation}
where  the sum is over the $C_\alpha$ for all finer-grained histories $\alpha$  contained within $\bar\alpha$.

For any individual history $\alpha$, there is a {\it branch state vector}  defined by
\begin{equation}
|\Psi_\alpha\rangle = C_\alpha |\Psi\rangle\, .
\label{twofive}
\end{equation}
When probabilities can be consistently assigned to the individual histories in a set,
they are given by
\begin{equation}
p(\alpha) = \parallel |\Psi_\alpha\rangle\parallel^2 = 
\parallel C_\alpha |\Psi\rangle\parallel^2\, .
\label{twosix}
\end{equation}
However, because of quantum interference,  probabilities cannot be consistently assigned to every set of alternative
histories that may be described.  The two-slit
experiment in Figure 1 provides an elementary example: An electron emitted by a source can pass
through either of two slits on its way to detection at a farther screen.  It would be
inconsistent to assign probabilities to the two histories distinguished by which slit the
electron goes through if no ``measurement'' process determines this.  Because of interference, the probability for arrival  at a point
on the screen would not be the sum of the probabilities to arrive there by going
through each of the slits. In quantum theory, probabilities are squares of amplitudes
and the square of a sum is not generally the sum of the squares.

Negligible interference between the branches of a set
\begin{equation}
\langle\Psi_\alpha|\Psi_\beta\rangle\approx 0 \quad , \quad \alpha\not=\beta \, ,
\label{twoseven}
\end{equation}
is a sufficient condition for the probabilities \eqref{twosix} to be consistent with the
rules of probability theory. The orthogonality of the branches is approximate in
realistic situations. But we mean by \eqref{twoseven} equality to an accuracy that defines probabilities well beyond the
standard to which they can be checked or, indeed, the physical situation modeled
\cite{Har91a}. 

Specifically, as a consequence of \eqref{twoseven}, the probabilities \eqref{twosix} obey
the most general form of the probability sum rules
\begin{equation}
p(\bar\alpha) \approx \sum_{\alpha\in\bar\alpha} p(\alpha)
\label{twoeight}
\end{equation}
for any coarse graining $\{{\bar\alpha}\}$ of the $\{\alpha\}$.  Sets of
histories obeying \eqref{twoseven} are said to (medium) decohere. As L.~Di\'osi has shown \cite{Dio04}, medium
decoherence is the weakest of known conditions that are consistent with elementary notions of the
independence of isolated systems\footnote{For a
discussion of the linear positive, weak, medium, and strong decoherence conditions,
see \cite{GH90b, GH95, Har04}.}. 
Medium-decoherent sets are thus the ones for which quantum mechanics makes predictions of consistent  probabilities through 
\eqref{twosix}.  The decoherent sets exhibited by our universe are determined through \eqref{twoseven} and by the Hamiltonian $H$ and the quantum state $|\Psi\rangle$.
The term {\it realm} is used as a synonym for a decoherent set of coarse-grained
alternative histories.

An important mechanism of decoherence is the dissipation of phase coherence between
branches into variables not followed by the coarse graining.  Consider by way of
example, a dust grain in a superposition of two positions deep in interstellar space
\cite{JZ85}.  In our universe, about $10^{11}$ cosmic background photons scatter from
the dust grain each second.  The two positions of the grain become correlated with different, nearly
orthogonal states of the photons. Coarse grainings that follow only the position of the
dust grain at a few times therefore correspond to branch state vectors that are nearly
orthogonal and satisfy \eqref{twoeight}.  

Measurements and observers play no fundamental role in this general formulation
of usual quantum
theory.  The probabilities of measured outcomes can be computed and are given to
an excellent approximation by the usual story. But, in a set of histories where they
decohere, probabilities can be assigned to the position of the Moon when it is not
receiving the attention of observers and to the values of density fluctuations in the early 
universe when there were neither measurements taking place nor observers to carry them out.

\end{document}